\documentstyle[12pt]{article}

\begin{document}
\title{{The Nuclear Effects in polarized proton-deuteron Drell-Yan processes }
}
\author{Duan Chun-Gui $^{1,2,4}$
 Shi Li-Jie  $^3$  Li Guang-Lie$^{2,4}$ Shen Peng-Nian$^{2,4}$
 \\
{\small 1.Department of Physics, Heibei Normal University, Shijiazhuang ,050016}\\
{\small 2.Institute of high energy physics,The Chinese academy of sciences,Beijing,100039}\\
{\small 3.Institute of  physics,The Chinese academy of
sciences,Beijing,100080}\\
{\small
4.CCAST(WorldLaboratory).P.O.Box8730,Beijing,100080,China}}
\date{}
\maketitle \baselineskip 9mm \vskip 0.5cm
\begin{center}
\begin{minipage}{120mm}
\begin{center}
Abstract
\end{center}
{\sl The longitudinally polarized Drell-Yan process is one of most
powerful tools to probe the structure of hadrons. By means of the
recent formalism of the polarized proton-deuteron(pd) Drell-Yan
,we calculate the ratio of the proton-deuteron Drell-Yan cross
section to the proton-proton(pp) one
$\Delta\sigma_{pd}/2\Delta\sigma_{pp}$in the polarized case.The
theoretical results can be compared with future experimental data
to confirm the nuclear effect due to the six-quark cluster in
deuteron .}


\end{minipage}
\end{center}
\vskip 0.5cm
\begin{flushleft}{\large\bf$ 1.$ Introduction}\end{flushleft}

One of the most active areas of research in nuclear and particle
physics during last several decades is the study of quark and
gluon distributions in the nucleons and nuclei. Several major
surprises were discovered in Deep-Inelastic Scattering (DIS)
experiments which profoundly changed our views of the partonic
substructure of hadrons $^{[1]}$. The unpolarized DIS can give
accurate measurements of structure functions $F_1(x,Q^2)$ and
$F_2(x,Q^2)$, which depend on x, the fractional momentum carried
by the struck parton, and $Q^2$, the four-momentum transfer
squared of the exchanged virtual photon. In the early 1980's the
European Muon Collaboration (EMC)$^{[2]}$ found in muon DIS
provided the first unambiguous evidence that the structure
functions in nuclei are significantly different from those in free
nucleons. This phenomenon is well-known as the EMC effect; which
implies that the quark distributions are different for a bound and
a free nucleon.

The $F_1$ and $F_2$ structure functions are sensitive to the
helicity-averaged parton distributions . Recent improvements in
polarized lepton beams and targets have made it possible to make
increasingly accurate measurements of two additional structure
functions $g_{1}(x,Q^{2})$ and $g_{2}(x,Q^{2})$, which depend on
the difference in parton  distributions with helicity either
aligned or anti-aligned with the spin of the parent
particles$^{[3]}$. In the naive quark-parton model $(QPM)$, the
nucleon is composed of quarks which have no orbital angular
momentum, and there are no polarized gluons present. In this
simple picture, the unpolarized structure function
$F_{1}(x,Q^{2})$ and polarized structure function $g_{1}(x,Q^{2})$
can be simply expressed as the charge weighed sum and difference
between momentum distributions for quark helicities aligned
paralleled $(q^{\uparrow})$ and anti-paralleled $(q^{\downarrow})$
to the longitudinally polarized nucleon:
\begin{eqnarray}
F_{1}(x,Q^{2})=\frac{1}{2}\sum_{i}e_{i}^{2}[q_{i}^{\uparrow}
(x,Q^{2})+q_{i}^{\downarrow}(x,Q^{2})],
\end{eqnarray}
\begin{eqnarray}
g_{1}(x,Q^{2})=\frac{1}{2}\sum_{i}e_{i}^{2}[q_{i}^{\uparrow}
(x,Q^{2})-q_{i}^{\downarrow}(x,Q^{2})]
\equiv\frac{1}{2}\sum_{i}e_{i}\Delta q_{i}(x,Q^{2}).
\end{eqnarray}
The charge of quark flavor u, d, and s is denoted by $e_{i}$, and
$q_{i}^{\uparrow(\downarrow)}(x,Q^{2})$ are the quark plus
antiquark momentum distribution.

In 1987, the EMC reported results from a polarized muon-proton
scattering experiment at CERN which puzzled the particle and
nuclear physics communities. Contrary to the prediction of the
naive quark model, the EMC found the little of the proton spin
seemed to be carried by the spins of the quarks (the so-called
"spin crisis")$^{[4]}$. Subsequent precision measurements are
consistent with the original experimental results, but the
theoretical interpretation has become more complex. It is now
believed that in additional to the quarks, the orbital angular
momentum and gluons may contribute significantly to the proton's
spin. The polarized structure functions are interesting not only
in opening a new degree of freedom with which to explore the
detailed structure of the nucleon, but also for making a precise
test of QCD via Bjorken sum rule which is a strict QCD
prediction$^{[5]}$.

The production of lepton pairs in hadron collisions, the Drell-Yan
process $^{[6]}$ is also  one of most powerful tools to probe the
structure of the nucleons and nuclei. Its parton model
interpretation is straightforward --- the process is induced by
the annihilation of a quark-antiquark pair into a virtual photon
which subsequently decays into a lepton pair. The Drell-Yan
process in proton-proton or proton-nucleus collisions therefore
provides a direct probe of the quark distribution in the nucleon
and nuclei. It is further natural to expect that a measurement of
the Drell-Yan cross section in polarized proton-nucleon(nuclei)
collision will yield information on the polarized quark
distribution in the nucleons (nuclei), which is an alternative
method to the DIS.

It has been commonly considered that the nuclear effects in
deuteron are neglected, and its structure functions is regarded as
the sum of the structure functions of the proton and neutron.
However, Gomez et. al $^{[7]}$ found that the deuteron has a
significant EMC effect. In addition, the E665 experimental result
$^{[8]}$ suggests the presence of nuclear shadowing effects in
deuteron. The analysis by Epele et al$^{[9]}$ also shows a
significant nuclear effects due to the composite nature of the
deuteron.

 Since the experimental discovery of the EMC effect, many
theoretical models have been put forward to explain it$^{[10]}$.
 The work by Lassila and Sukhatme$^{[11]}$ shows that the quark
  cluster model (QCM) can give a good unified
explanation for the experimental data of the nuclear effects in
whole x region. Spin-dependent effects in the QCM of the deuteron
and $^{3}He$ were investigated by Benesh and Vary$^{[12]}$. But
there were no detailed 6-quark clusters quark distribution by
them. Several years ago, Brodsky, Burkardt and Schmidt provided a
reasonable description of the spin-dependent quark distributions
of the nucleon in a pQCD based model$^{[13]}$. This analysis have
been  extended  to the description of the spin-dependent quark
distributions in a 6-quark cluster$^{[14]}$. In previous
work$^{[15]}$,by means of the polarized quark distributions in a
6-quark cluster, the  nuclear effects on polarized structure
function in deuteron  have been investigated. It is found that the
calculated results with nuclear effects can better fit the SLAC
E155 experimental data$^{[16]}$ than that without nuclear effects
.In order to further investigate the nuclear effects in  deuteron,
an alternative way is given  by combining the pd Drell-Yan data
with the pp data in the polarized case.

\begin{flushleft}{\large\bf$ 2.$ Theoretical Scheme}\end{flushleft}

Recently,the formalism of the polarized pd Drell-Yan processes had
been available. A theoretical formalism had been completed for the
longitudinally polarized pd Drell-Yan processes$^{[17,18]}$.
Taking advantage of the formalism, the nuclear-effects in deuteron
can be discussed by measuring the ratio of the polarized pd
Drell-Yan cross section to the polarized  pp one
$\Delta\sigma_{pd}/2\Delta\sigma_{pp}$. In the Ref.[17], the
difference between the longitudinally-polarized pd cross section
is given by
\begin {eqnarray}
\Delta\sigma_{pd}=\sigma(\uparrow_L,-1_L)-\sigma(\uparrow_L,+1_L)
\propto-\frac{1}{4}[2V^{LL}_{0,0}
+(\frac{1}{3}-\cos^2\theta)V^{LL}_{2,0}],
\end {eqnarray}
where the subscripts of  $\uparrow_L, +1_L,$ and $-1_L$ indicate
the longitudinal polarization , $\sigma({\sl pol}_p,{\sl pol}_d)$
indicates the cross section with the proton polarization $pol_p$
and the deuteron one $pol_d$. The longitudinally polarized
structure functions $V^{LL}_{0,0}$ and $V^{LL}_{2,0}$ are defined
in Ref.[17]. The subscripts ${\sl l}$ and ${\sl m}$ of the
expression $V^{LL}_{{\sl l,m}}$ indicate that it is obtained by
the integration $\int d \Omega Y_{lm}\Delta\sigma_{pd}$, and the
superscript LL means that proton and deuteron are both
longitudinally polarized. The $\theta$ is the polar angle of the
final lepton $\mu^+$. A parton model should be used for discussing
relations between the structure functions and polarized parton
distributions. In the following, we employ the expression which is
obtained by integrating the cross section over the virtual-photon
transverse momentum $Q_T$. According to Ref.[18], it is given by
\begin {eqnarray}
\Delta\sigma_{pd}\propto \sum\limits_f  e^2_f [\Delta q_f
(x_1)\Delta\bar{q}^d_f (x_2)+\Delta\bar{q}_f (x_1)\Delta q^d_f
(x_2)],
\end {eqnarray}
where $\Delta q_f (\Delta q^d_f )$ and $\Delta\bar{q}_f
(\Delta\bar{q}^d_f )$ are the longitudinally-polarized quark and
antiquark distributions function in the proton(deuteron). The
subscript $f $ indicates quark flavor, and $e_f $ is the
corresponding quark charge.$x_1$ and $x_2$ are the momentum
fractions of proton (deuteron ) carried by the quark or anti-quark
.

If  we disregard the nuclear effects in  deuteron and assume
isospin symmetry, the polarized quark distribution functions in
deuteron can be expressed as $$\Delta u^d=\Delta u+\Delta d,
\Delta d^d=\Delta d+\Delta u, \Delta s^d=2\Delta s,$$
\begin {eqnarray}
 \Delta \bar{u}^d=\Delta \bar{u}+\Delta \bar{d},
\Delta \bar{d}^d=\Delta \bar{d}+\Delta \bar{u}, \Delta
\bar{s}^d=2\Delta \bar{s},
\end {eqnarray}
where $\Delta q$ ( $\Delta \bar{q}$) is quark ( anti-quark)
polarization distribution function in proton. Similarly, the pp
Drell-Yan  cross section are given by simply substituting
$q^d(\bar{q}^d)$ with $q(\bar{q})$. Therefore ,the ratio of the pd
cross section to the pp one is then obtained as,
\begin {eqnarray}
R_{pd}=\frac{\Delta\sigma_{pd}}{2\Delta\sigma_{pp}}=\frac{\sum\limits_
f  e^2_f [\Delta q_f (x_1)\Delta\bar{q}^d_f (x_2)+ \Delta\bar{q}_f
(x_1)\Delta q^d_f (x_2)]}{2\sum\limits_f  e^2_f [\Delta q_f
(x_1)\Delta\bar{q}_f (x_2)+ \Delta\bar{q}_f (x_1)\Delta q_f
(x_2)]}
\end {eqnarray}

The behavior of $R_{pd}$ at  two $x_F$ extreme limits have been
analyzed by Kumano and Miyama$^{[19]}$. If two extreme limits
($x_F=x_1-x_2\rightarrow\pm 1)$ are taken in Eq.(6) with the
assumption $\Delta u_v(x\rightarrow 1)\gg \Delta d_v(x\rightarrow
1$)$^{[13,20]}$, the ratio becomes
\begin {eqnarray}
R_{pd}(x_F\rightarrow+1)=\frac{1}{2}[1+\frac{\Delta\bar{d}(x_2)}{\Delta\bar{u}(x_2)}]_{x_2\rightarrow0}
\end {eqnarray}
\begin {eqnarray}
R_{pd}(x_F\rightarrow-1)=\frac{1}{2}[1+\frac{\Delta\bar{d}(x_1)}{4\Delta\bar{u}(x_1)}]_{x_1\rightarrow0}
\end {eqnarray}
It is found that the flavor-asymmetric distribution $\Delta
\bar{u}-\Delta \bar{d}$ can be extracted by finding the deviation
from 1 at $x_F\rightarrow\pm1$ or from $5/8$ at
$x_F\rightarrow-1$. However, $R_{pd}$ in other $x_F$ regions are
not so promising in the flavor asymmetry, and can be used to find
the x dependence of the polarized quark distributions function in
deuteron  so that the nuclear effects in  deuteron  can be shed
light on .

Now let's turn to investigate the nuclear Effects in polarized
proton-deuteron Drell-Yan processes. In the quark cluster model,
the presence of 6-quark cluster is used to understand the nuclear
effects in  deuteron. Therefore, when we take account of the
nuclear  effects in  deuteron and employ isospin symmetry, the
polarized quark distribution function in deuteron can be written
as
$$\Delta u^d=p_3(\Delta u+\Delta d)+p_6\Delta u^6,
\Delta d^d=p_3(\Delta d+\Delta u)+p_6\Delta d^6,$$
$$\Delta s^d=2p_3\Delta s+p_6\Delta s^6,\Delta \bar{u}^d=p_3(\Delta \bar{u}+\Delta
\bar{d})+p_6\Delta\bar{u}^6,$$
\begin {eqnarray}
 \Delta\bar{d}^d=p_3(\Delta
\bar{d}+\Delta \bar{u})+p_6\Delta \bar{d}^6,
\Delta\bar{s}^d=2p_3\Delta \bar{s}+p_6\Delta\bar{s}^6,
\end {eqnarray}
 where$\Delta q^6$ ( $\Delta \bar{q}^6$) is quark ( anti-quark)
polarization distribution function in deuteron ,
$p_3=[(p_{s}-p_{6s})-\frac{1}{2}(p_{d}-p_{6d})]$ is the
possibilities for creating $3$-quark cluster , $p_{s}=0.957$ and
$p_{d}=0.043$ denote the probabilities for finding the deuteron in
an s or d wave, respectively. $p_{6}=p_{6s}+p_{6d}$, in which
$p_{6s}=0.047$ and $p_{6d}=0.007$ calculated by Benesh and
Bary$^{[12]}$ are the probabilities for creating a 6-quark cluster
in the s- and d-states , denote the possibilities for creating
$6$-quark cluster in deuteron .

\begin{flushleft}{\large\bf$ 3.$ Results and Discussions}\end{flushleft}

With all ingredient sets as given above, we can numerically
calculate the ratio of the polarized pd Drell-Yan cross section to
the pp one $\Delta\sigma_{pd}/2\Delta\sigma_{pp}$with nuclear
effects and without nuclear effects in deuteron. In our
calculation, the
 polarized parton distribution functions for proton  are taken from the
 new version of the LSS(Lead-Sidorov-Stamenev) leading-order(LO)
parameterization$^{[21]}$. In Fig.1,the theoretical results are
given with taking  center-of-mass energy $\sqrt{s}=50 $GeV and
dimuon mass $M_{\mu\mu}=5$Gev.
 The solid curve denotes the $R_{pd}$ with the nuclear effects  due to
 6-quark clusters in deuteron , the dashed curve corresponds to not considering
 the 6-quark clusters. In addition,the Drell-Yan cross section  ratio $R_{pd}$ is
 calculated at  $\sqrt{s}=200$GeV  and $M_{\mu\mu}=5$GeV ,the results are shown in Fig.2.
It is obvious that the nuclear effects  in deuteron become more
significant at higher  center-of-mass energy and large $x_F$.
Because the difference between $R_{pd}$with  nuclear effects and
without nuclear effects in the range $x_F>0.08$ is larger  at
higher center-of-mass energy, this makes it possible to further
confirm the nuclear effects in deuteron .

In summary , we have investigated the longitudinally polarized
Drell-Yan cross section ratio
$\Delta\sigma_{pd}/2\Delta\sigma_{pp}$ by means of the recent
formalism for the polarized pd Drell-Yan processes . It is  shown
that the information on the nuclear effects in deuteron can be
extracted in the future experiment. Although there is not
experiments for the polarized pd Drell-Yan   at this stage , we
suggest precise experimental research on this reaction at FNAL
,HEAR ,and RICH ,which makes us good understanding the unclear
effects in deuteron .

{\bf Acknowledgement:}This work is partially supported by  Major
State Basic Research Development program (under Contract No.
G20000774), by the National Natural Science Foundation of
China(No.19875024,19835010,10175074), and by Natural Science
Foundation of Hebei Province(No.103143)

\vskip 1cm
\begin{center}
Figure caption
\end{center}
Fig.1 The Drell-Yan cross section  ratio $R_{pd}$ with $\sqrt{s
}=50$GeV  and $M_{\mu\mu}=5$GeV . The solid curve corresponds to
including contributions of the 6-quark clusters, i. e. nuclear
effects . The dashed curve denotes the results without 6-quark
clusters in deuteron.\\
Fig.2 The Drell-Yan cross section  ratio $R_{pd}$ with $\sqrt{s
}=200$GeV  and $M_{\mu\mu}=5$GeV . The comments are the same as
Fig. 1.

\end{document}